\documentstyle[preprint,aps]{revtex}
\begin{document}
\draft
\title{Equivalence between different classical 
treatments of the $O(N)$ non-linear sigma model and 
their functional Schr\"{o}dinger equations.} 
\author{A. A. Deriglazov\thanks{alexei@fisica.ufjf.br},
W. Oliveira\thanks{wilson@fisica.ufjf.br}  and 
G. Oliveira-Neto\thanks{gilneto@fisica.ufjf.br}} 
\address{Departamento de F\'{\i}sica,
Instituto de Ciencias Exatas,
Universidade Federal de Juiz de Fora,
CEP 36036-330, Juiz de Fora,
Minas Gerais, Brazil.}
\date{\today}
\maketitle

\begin{abstract}
In this work we derive the Hamiltonian formalism of the 
$O(N)$ non-linear sigma model in its original version as 
a second-class constrained field theory and then as a 
first-class constrained field theory. We treat the model 
as a second-class constrained field theory by two 
different methods: the unconstrained and the Dirac 
second-class formalisms. We show that the Hamiltonians for 
all these versions of the model are equivalent. Then, for a 
particular factor-ordering choice, we write the functional  
Schr\"{o}dinger equation for each derived Hamiltonian. 
We show that they are all identical which justifies our
factor-ordering choice and opens the way for a future
quantization of the model via the functional 
Schr\"{o}dinger representation.
\end{abstract}

\section{Introduction }
\label{sec:one}

The Hamiltonian formulation of a classical constrained 
system with second-class constraints can be obtained, 
usually, in two different ways \cite{sudarshan}. In the 
first one, the unconstrained formalism, one starts solving 
the classical constraints and substituting the result in 
the action. Then, one proceeds with the derivation of the 
Hamiltonian in the usual way because the theory is written, 
now, in terms of the physical degrees of freedom only. In 
the second one, the Dirac second-class, one
writes the Hamiltonian formalism of the theory with the
Dirac brackets which take in account the constraints 
explicitly. After that, the constraints can be solved and
part of the variables can be omitted from consideration. In 
general, the remaining physical variables are not
canonically conjugated to their momenta through the Dirac
bracket. Then, one finds transformations
to new variables which are canonically conjugated to their
momenta through the Dirac bracket. Finally, one writes 
the reduced Hamiltonian, obtained after the use of the 
relations derived from the constraints in the original 
Hamiltonian.

For some time now, other methods of treating a 
second-class constrained system have been developed 
\cite{BFF,BT,clovis,wilson}. One converts the original 
theory in a theory with first-class constraints and 
derives its Hamiltonian. The main motivation for this 
conversion are the symmetries that the first-class systems 
possess. Through the symmetries, it is possible to 
determine many physical properties of the system in a more 
general way. Therefore, one expects to use those symmetries 
to study the properties of a second-class system after the 
conversion to a first-class one. 

Although one expects that the different ways to treat a 
second-class constrained system leads to the same Hamiltonian
theory, it is by no means trivial to show explicitly. In the
present work we would like to show this equivalence for
the $O(N)$ non-linear sigma model, which is a well-know
second-class constrained field theory \cite{novikov,abdalla}.

We shall consider the $O(N)$ non-linear sigma model described
in a $1+1$-dimensional Minkowski space-time. Therefore, it
cannot directly describe physical phenomena in the real world.
On the other hand, since it was shown to have many properties
similar to physically relevant, $3+1$-dimensional, non-Abelian
field theories \cite{novikov,abdalla,novikov1}, we believe 
that our result will be easily extended to more physically 
relevant theories. Over the years, many works have been
dedicated to the quantization of the $O(N)$ non-linear sigma
model using different techniques 
\cite{abdalla,maharana,bardeen,davis,banerjee,kim}. As we 
shall see below, some of the works dealing with the canonical 
quantization of the model have few points of contact with our 
work.

Besides the purely classical treatment, we shall also 
write down the functional Schr\"{o}dinger equation 
\cite{hatfield,jackiw} for each derived Hamiltonian. 
As we shall see, for a particular factor-ordering 
choice, the functional Schr\"{o}dinger equations are 
identical. This result, along with the fact that
in the study of the first class constrained version of the 
model the ordering is consistent with the operatorial 
version of the classical constraint algebra (see Sec. 
\ref{sec:three}), justifies our factor-ordering
choice and opens the way for a future quantization of the
model via the functional Schr\"{o}dinger representation.
It is important to notice that, we shall not demonstrate
that our factor-ordering choice is the only one to 
satisfy the above mentioned properties. It means that, 
there may be other choices that also satisfy those 
properties.

The functional Schr\"{o}dinger representation
has recently been systematically used in order
to quantize different field theories, including
gravity \cite{hatfield,jackiw,jackiw1,gil}. 
Many theoretical as
well as some physical predictions have been
derived, for different theories, from the
wave-functionals obtained so far. One example of
an important theoretical feature of gauge 
theories established in the context of the
functional Schr\"{o}dinger representation,
without any `instanton' approximation, is the
so-called vacuum angle \cite{jackiw}. On the other
hand, from the wave-functional of the quantum
Schwarzschild-de Sitter black hole one is able
to predict how it depends on the mass and
cosmological constants \cite{gil}.

In the next section, Sec. \ref{sec:two}, we shall treat 
the $O(N)$ non-linear sigma model as a second-class 
constrained field theory. We shall use the unconstrained and 
the Dirac second-class formalisms, both described above. As 
we shall see, the resulting Hamiltonians coming from both 
formalisms, written in terms of the initial fields, will be 
the same. Therefore, they are classically equivalent. It 
also means that, the two formalisms will lead to the same 
functional Schr\"{o}dinger equation, if we apply the same
factor-ordering choice for each of them. Here, besides the main 
result, further novelties in relation to previous works in
this area will appear. In the treatment of the model using the
unconstrained formalism, we shall not use the standard field
transformation in order to eliminate one of the fields
\cite{bardeen,davis,kim}. Rather, we shall express one of the 
fields in terms of the others through the constraint. Therefore, 
our Lagrangian present in the action eq. (\ref{5}) and 
Hamiltonian eq. (\ref{12}) will be different from the ones in 
\cite{bardeen,davis,kim}. Since, in Ref. \cite{kim} they also 
work in the functional Schr\"{o}dinger 
representation, the difference in the Hamiltonians implies that 
the functional Schr\"{o}dinger equations will not be the same.
Nevertheless, due to the fact that the difference comes from
the use of different field basis in order to describe the 
model we should obtain the same results from our Hamiltonian
as the ones found in Ref. \cite{kim}. In a recent work
on the analogous quantum mechanical problem of a particle
moving on a sphere the authors used the unconstrained formalism
in order to write the Hamiltonian of that system  \cite{banerjee1}.
Then, if we keep in mind the differences between the two systems
our Lagrangian and Hamiltonian will be similar to theirs.
In the treatment of the 
model with the Dirac second-class formalism, we shall 
explicitly write transformations eq. (\ref{28}) 
from the original fields and conjugated momenta to new ones, 
such that, the Dirac brackets between the new fields and their 
conjugated momenta have the canonical form. These 
transformations eq. (\ref{28}) and consequently the 
new fields and their conjugated momenta are different from the
ones introduced in previous works \cite{maharana,davis}.

In Sec. \ref{sec:three}, we shall treat the $O(N)$ non-linear
sigma model as a first-class constrained field theory. The
Hamiltonian that we shall manipulate was first derived in 
\cite{wilson}. We shall show that this Hamiltonian leads to
the same classical theory than the other two Hamiltonians 
obtained in Sec. \ref{sec:two}. Then, we shall derive the
functional Schr\"{o}dinger equation using the Dirac 
first-class quantization technique \cite{dirac}. There, one
writes the operatorial versions of the constraints and forces
them to annihilate the wave-functional. Then, this 
wave-functional that satisfies the operatorial version of the
constraints, must be a solution to the functional 
Schr\"{o}dinger equation. In the present case, as we shall 
see, after we use the information coming from the annihilation 
of the wave-functional by the quantum constraints, the 
functional Schr\"{o}dinger equation reduces to the one 
obtained in Sec. \ref{sec:two}. 

Finally, in Sec. \ref{sec:four} we summarize the main points 
and results of the paper.

\section{The $O(N)$ non-linear sigma model as a second-class
constrained field theory.}
\label{sec:two}

The $O(N)$ nonlinear sigma model is described by the action,

\begin{equation}
S =\, \int d^2x\, \left(\, {1\over 2}\, \partial_\mu\phi^a
\partial^\mu\phi^a\, \right)\, ,
\label{1}
\end{equation}
where it is implied the kinematic constraint,
\begin{equation}
\label{2}
T_1\, =\, |\phi|^2\, -\, 1\, .
\end{equation}
Here, $\mu=0,1$, $a$ is an index related to the $O(N)$
symmetry group, the metric has signature (+,-), 
we are using the convention of sum over repeated index
and $|\phi|^2\equiv \phi^a\phi^a$.

In the present section we shall treat the $O(N)$ non-linear 
sigma model as a second-class constrained field theory. We
shall obtain its Hamiltonian formulation in two different ways 
\cite{sudarshan}. In the first one, Subsec. \ref{subsec:one}
we shall start expressing one of the fields in terms of the
others, through the constraint, and substituting the result 
in the Lagrangian. Next, we shall find the Hamiltonian. The 
theory will be written, then, in terms of the physical 
degrees of freedom only. Finally, using the Hamiltonian and
a particular factor-ordering choice we shall write the
functional Schr\"{o}dinger equation. As explained above, 
the Lagrangian present in the action eq. (\ref{5}) and the 
Hamiltonian eq. (\ref{12}) will be different from the ones 
in \cite{bardeen,davis,kim} and similar to the ones in
\cite{banerjee1}, if we keep in mind the differences between
our model and the one in \cite{banerjee1}. In the second way, 
Subsec. 
\ref{subsec:two} we shall write the Hamiltonian formalism of 
the theory with the Dirac brackets which take in account the 
constraints explicitly, following \cite{maharana,davis}. As 
we shall see, the initial variables are not canonically 
conjugated to their momenta through the Dirac brackets. Here, 
a novelty with respect to the treatments of Refs. 
\cite{maharana,davis} will appear: we shall explicitly 
introduce transformations eq. (\ref{28}) to new 
variables which are canonically conjugated to their momenta 
through the Dirac brackets. Then, we shall impose the 
constraints and write the reduced Hamiltonian, obtained after 
the use of the relations derived from the constraints in the 
original Hamiltonian. Finally, we re-write the reduced
Hamiltonian in terms of the new variables. We shall call it
physical Hamiltonian ($H_{phys}$). As we shall see this 
physical Hamiltonian is identical to the one obtained in 
Subsec. \ref{subsec:one}. Naturally, if one uses the same 
factor-ordering choice of Subsec. \ref{subsec:one}, the 
functional Schr\"{o}dinger equation derived from that physical 
Hamiltonian must be the same as the one computed in Subsec.
\ref{subsec:one}.

\subsection{Unconstrained formalism.}
\label{subsec:one}

We start by strongly imposing the constraint $T_1$ eq. 
(\ref{2}). Then, we write one of the fields, say $\phi^N$,
in terms of the other $N-1$ fields,

\begin{equation}
\label{3}
\phi^N\, =\, \sqrt{1 - \phi^i\phi^i}\, ,
\end{equation}
where $i=1,2,...,N-1$.
From eq. (\ref{3}) it is straightforward to compute 
$\partial_\mu \phi^N$ as,

\begin{equation}
\label{4}
\partial_\mu \phi^N\, =\, - {\phi^i \partial_\mu \phi^i
\over \sqrt{1 - \phi^i\phi^i}}\, .
\end{equation}

Introducing both results eqs. (\ref{3}) and (\ref{4}) in
the action eq. (\ref{1}), we obtain the theory described in
terms of the $N-1$ physical fields.

\begin{equation}
\label{5}
S_{phys}\, =\, \int d^2x \left(\, {1\over 2}\, g_{ij}
\partial_\mu \phi^i \partial_\mu \phi^j\, \right)\, ,
\end{equation}
where $g_{ij}$ is given by,
\begin{equation}
\label{6}
g_{ij}\, =\, \delta_{ij}\, +\, {\phi_i \phi_j\over 1 - 
\phi^i\phi^i}\, .
\end{equation}

Now, we would like to construct the Hamiltonian of the
model for posterior quantization. The initial step is
the derivation of the momenta, through the usual definition,

\begin{equation}
\label{7}
\pi_i\, =\, {\partial L\over \partial (\partial_0 
{\phi}^i)}\, ,
\end{equation}
where $L$ is the density of Lagrangian which can be read
directly from $S$ eq. (\ref{5}) and $\partial_0$ means
partial derivative with respect to the time coordinate.
So, we may compute the momenta to obtain,

\begin{equation}
\label{8}
\pi_i\, =\, g_{ij}\partial_0 \phi^j.
\end{equation}

In order to re-write the theory in its Hamiltonian form
we must know how to invert eq. (\ref{8}), so that, we may
write the {\it velocities} in terms of the momenta. It is
accomplished by the computation of the inverse of $g_{ij}$
which is,

\begin{equation}
\label{9}
\tilde{g}^{ij}\, =\, \delta^{ij}\, -\, \phi^i \phi^j\, .
\end{equation}
Therefore,

\begin{equation}
\label{10}
\partial_0\phi^i\, =\, \tilde{g}^{ij} \pi_j\, .
\end{equation}

The Hamiltonian of the theory, which general expression is,

\begin{equation}
\label{11}
H\, =\, \int dx ( \pi_i \partial_0 \phi^i\, -\, L )\, ,
\end{equation}
takes the particular form,

\begin{equation}
\label{12}
H\, =\, \int dx \left( \, {1\over 2}\, \tilde{g}^{ij} 
\pi_i \pi_j\, +\, {1\over 2}\, g_{ij} \partial_x \phi^i 
\partial_x \phi^j \, \right)\, .
\end{equation}
Where $\partial_x$ means
partial derivative with respect to the spatial coordinate.

By definition the ($\phi^i$, $\pi_i$) form canonically 
conjugated pairs which have the usual Poisson brackets,

\begin{equation}
\label{13}
\{\phi^i(x_0,x) , \pi_j(x_0,x')\}\, =\, \delta^i_j 
\delta^{N-1}(x-x')\, .
\end{equation}

As we have mentioned above, the unconstrained formalism
was recently used in the study of the analogous quantum
mechanical problem of a particle moving on a sphere
\cite{banerjee1}. Therefore, if we keep in mind the
differences between the two systems, some of the above
equations (\ref{3}-\ref{12}) are similar to theirs.

Now, we would like to write 
the functional Schr\"{o}dinger equation of the model
\cite{hatfield,jackiw}. Therefore, we 
start introducing the wave-functional $\Psi[\phi^i,t]$. 
Then, we consider the $\phi^i$'s and the $\pi_i$'s 
as quantum operators, it means that in the fields 
representation the momenta are replaced by the following 
functional derivatives,

\begin{equation}
\label{14}
\pi_i (x)\, \to\, - i { \delta \over \delta \phi^i (x)}\, ,
\end{equation}
where we have set $\hbar$ equal to one.

The wave-functional $\Psi$ satisfies the functional 
Schr\"{o}dinger equation,

\begin{equation}
\label{15}
i {\partial \over \partial t} \Psi [\phi^i, t]\,
=\, \hat{H} [\phi^i, t] \Psi [\phi^i, t]\, ,
\end{equation}
where $\hat{H}$ is the operatorial version of $H$ eq.
(\ref{12}).

It is important to notice that since $\tilde{g}^{ij}$
depends on the fields, the kinetic term in the
Hamiltonian eq. (\ref{12}) will develop factor-ordering
ambiguities upon quantization. Here, we shall solve this 
problem by choosing a particular factor-ordering. We 
shall write all field functions to the left of the
momenta operators. We justify this choice by two different facts.
Firstly, its application in the Hamiltonians
obtained in the present paper leads to the same functional
Schr\"{o}dinger equation. Secondly, in the study of
the first class constrained version of the model
the ordering is consistent with the operatorial version 
of the classical constraint algebra (see Sec. \ref{sec:three}). 
The situation is similar to the one with the Wheeler-DeWitt 
equation
 \cite{isham}. As we have mentioned above, we shall
not demonstrate that our factor-ordering choice is the only 
one to satisfy the above mentioned properties. It means that, 
there may be other choices that also satisfy those properties.

Taking in account the explicit expression of $H$ eq. 
(\ref{12}) and the particular factor-ordering choice
mentioned above, the functional Schr\"{o}dinger equation 
for the O(N) non-linear sigma model is given by,

\begin{equation}
\label{16}
\int dx \left(\, {1\over 2}\, \tilde{g}^{ij}{\delta^2 \Psi 
\over \delta \phi^i \delta \phi^j}\, +\, {1\over 2}\, g_{ij}
\partial_x \phi^i \partial_x \phi^j\Psi \, \right)\, =\,
i {\partial \over \partial t} \Psi\, .
\end{equation}

Since the Hamiltonian eq. (\ref{12}) does not explicitly
depend on time, we may separate out the time dependence of 
the wave-functional and write,

\begin{equation}
\label{17}
\Psi[\phi^i , t]\, =\, e^{-iEt} \Psi [\phi^i]\, .
\end{equation}

From eq. (\ref{16}), $\Psi [\phi^i ]$ satisfies the 
time-independent functional Schr\"{o}dinger equation,

\begin{equation}
\label{18}
\int dx \left(\, {1\over 2}\, \tilde{g}^{ij}{\delta^2 
\Psi \over \delta \phi^i \delta \phi^j}\, +\, {1\over 2}\, 
g_{ij} \partial_x \phi^i \partial_x \phi^j\Psi \, \right)\, 
=\, E \Psi\, .
\end{equation}

It is clear from the above equation (\ref{18}) that the
energies $E$, the eigenvalues of the Hamiltonian, will be
determined for the present model when we solve this 
equation.

\subsection{Dirac second-class formalism.}
\label{subsec:two}

In the present formulation, it is more appropriated to
write the action of the model in the following way,

\begin{equation}
S =\, \int d^2x \left[\, {1\over 2}\, \partial_\mu\phi^a
\partial^\mu\phi^a\, +\, \lambda ( |\phi|^2 - 1 )\, \right]
\, ,
\label{19}
\end{equation}
where $\lambda$ is a Lagrange's multiplier and the geometrical
constraint was introduced in the action. It is not difficult
to see that the Lagrange's equations for both actions eqs. 
(\ref{1}) and (\ref{19}) are the same.

In order to write the Hamiltonian for the action eq. (\ref{19}),
we compute the canonically conjugated momenta. They are
given by eq. (\ref{7}) which for the present case reduce to,

\begin{equation}
\label{20}
\pi_a\, =\, \partial_0 \phi^a\, , \qquad \pi_\lambda\, =\, 0\, .
\end{equation}

Using the values of the momenta eq. (\ref{20}) and the 
Lagrangian present in the action (\ref{19}), the 
Hamiltonian eq. (\ref{11}) becomes,

\begin{equation}
\label{21}
H\, =\, \int dx \left[\, {1\over 2}\, \pi_a \pi_a\, +\, 
{1\over 2}\, \partial_x \phi^a \partial_x \phi^a\, -\, \lambda 
(|\phi|^2 - 1)\, +\, v_\lambda \pi_\lambda\, \right]\, ,
\end{equation}
where $v_\lambda$ is a new Lagrange's multiplier associated to
the constraint $\pi_\lambda = 0$.

We may now, derive all second-class constraints of the model
by computing the time evolution of the known constraints. 
Starting with the known constraint $\pi_\lambda = 0$ we obtain
the complete set,

\begin{equation}
\label{22}
\pi_\lambda\, =\, 0\, , \qquad |\phi|^2\lambda\, +\, {1\over 2}
\pi_a \pi_a\, +\, \partial_x \phi^a \partial_x \phi^a\, =\, 0\, ,
\end{equation}
\begin{equation}
\label{23}
G_1\, =\, |\phi|^2\, -\, 1\, =\, 0\, \qquad G_2\, =\, \phi^a
\pi_a\, =\, 0\, .
\end{equation}

Following Dirac's procedure \cite{dirac}, we take in account 
the above constraints eqs. (\ref{22}) and (\ref{23}) by 
constructing the Dirac bracket. After that, we shall be able 
to use explicitly the constraints in the theory.

The sector of the Dirac bracket involving the constraints eq.
(\ref{22}) is trivial. Therefore, it will get contributions 
just from $G_1$ and $G_2$ eq. (\ref{23}),

\begin{equation}
\label{24}
\{ A , B\}_D\, =\, \{ A , B \}\, +\, {1\over 2 |\phi|^2} \{
A , |\phi|^2 - 1 \} \{ \phi^a \pi_a , B \}\, -\, {1\over 2
|\phi|^2} \{ A , \phi^a \pi_a \} \{ |\phi|^2 - 1 , B \}\, ,
\end{equation}
where $A$ and $B$ are functions of the canonical variables
and all brackets in the right hand side of eq. (\ref{24})
are Poisson brackets.

Computing the Dirac brackets of the fields and their
conjugated momenta we obtain the below values,

\begin{eqnarray}
\label{25}
\{ \phi^a (x), \phi^b (x') \}_D & = & 0\, ,\quad \{ \pi_a 
(x), \pi_b (x')\}_D\, =\, \left( {\pi_a \phi^b - \pi_b 
\phi^a \over |\phi|^2} \right) \delta (x-x')\, ,\nonumber \\
\{ \phi^a (x), \pi_b (x')\}_D & = & \left( \delta^a_b - {
\phi^a \phi^b \over |\phi|^2} \right) \delta (x-x')\, .
\end{eqnarray}

At this stage we may use explicitly the results coming from 
the constraints. From the constraints eq. (\ref{22}) we 
learn that $\pi_\lambda = 0$ and the value of $\lambda$ in 
term of the other variables. From $G_1$ eq. (\ref{23}) we 
obtain eq. (\ref{3}) and from $G_2$ eq. (\ref{23}) we obtain,

\begin{equation}
\label{26}
\pi_N\, =\, -\, { \phi^i \pi_i \over \sqrt{1 - 
\phi^i\phi^i}}\, .
\end{equation}

Now, the model can be written in terms of $N-1$ independent
fields and their conjugated momenta. Using the 
results coming from the constraints, the Dirac brackets for
the $N-1$ independent fields and their conjugated momenta 
become,

\begin{eqnarray}
\label{27}
\{ \phi^i (x), \phi^i (x') \}_D & = & 0\, ,\quad \{ \pi_i 
(x), \pi_j (x')\}_D\, =\, \left( \pi_i \phi^j - \pi_j 
\phi^i \right) \delta (x-x')\, ,\nonumber \\
\{ \phi^i (x), \pi_j (x')\}_D & = & \left( \delta^i_j - 
\phi^i \phi^j \right) \delta (x-x')\, .
\end{eqnarray}

Since $\phi^i$'s and $\pi_i$'s do not form canonically 
conjugated sets, the next step \cite{dirac,gitman} 
is to find new variables which form canonical pairs
with relation to the Dirac brackets eq. (\ref{27}). It
allows one to apply the standard quantization methods, 
in particular, to write the functional Schr\"{o}dinger 
equation for the theory.

The variables are given by the following transformations,

\begin{equation}
\label{28}
\tilde{\phi}^i\, =\, \phi^i\, ,\qquad \tilde{\pi}_i\, =\,
\pi_i\, +\, \left( {\pi_j \phi^j \over 1 - \phi^i\phi^i} 
\right)\,\phi^i\, .
\end{equation}
They are easily inverted 
resulting in,

\begin{equation}
\label{29}
\phi^i\, =\, \tilde{\phi}^i\, ,\qquad \pi_i\, =\,
\tilde{\pi}_i\, -\, \left( {\tilde{\pi}_j \tilde{\phi}^j 
\over 1 + \tilde{\phi}^i\tilde{\phi}^i}\right)\, 
\tilde{\phi}^i\, ,
\end{equation}

Finally, we are in position to write the physical
Hamiltonian ($H_{phys}$) starting from $H$ eq.
(\ref{21}). For this we start substituting the constraints
eqs. (\ref{22}) and (\ref{23}) in $H$ eq. (\ref{21}). Then, 
we eliminate the non-physical variable $\phi^N$ and its
conjugated momentum $\pi_N$ through the eqs. (\ref{3}) and
(\ref{26}). After that, we re-write the whole expression in
terms of the canonically conjugated pairs ($\tilde{\phi}^i$,
$\tilde{\pi}_i$) eq. (\ref{28}). We obtain,

\begin{equation}
\label{30}
H_{phys}\, =\, \int dx \left( {1\over 2} \tilde{g}^{ij} 
\tilde{\pi}_i \tilde{\pi}_j\, +\, {1\over 2} g_{ij} 
\partial_x \tilde{\phi}^i \partial_x \tilde{\phi}^j
\right)\, .
\end{equation}

Comparing $H_{phys}$ eq. (\ref{30}) with $H$ eq. (\ref{12})
we can see that they are the same\footnote{Note that, eq.
(\ref{28}) means that the variables ($\phi^i, \pi_i$) of
the theory eq. (\ref{19}) and the corresponding variables
of the unconstrained formulation eq. (\ref{12}) are related
by the noncanonical transformations eq. (\ref{28}). We are
grateful to the referee for pointing this fact.}. Naturally, 
if one uses 
the same factor
ordering choice of Subsec. \ref{subsec:one},
the functional Schr\"{o}dinger equations for both must be 
identical since they are written in terms of pairs of 
fields and momenta that are canonically conjugated.

\section{The $O(N)$ non-linear sigma model as a first-class
constrained field theory.}
\label{sec:three}

The Hamiltonian for the $O(N)$ non-linear sigma model written 
as a first-class constrained field theory that we shall use 
is \cite{wilson}, 

\begin{equation}
\label{31}
H\, =\, \int dx \left[\, {1\over 2}\, \bar{g}^{ab} \pi_a 
\pi_b + \,{1\over 2}\, \partial_x \phi^a \partial_x \phi^a\, 
-\, \lambda (|\phi|^2 -1) + v_\lambda \pi_\lambda\, \right]\, ,
\end{equation}
where,
\begin{equation}
\label{32}
\bar{g}^{ab}\, =\, \delta^{ab} - {\phi^a \phi^b\over |\phi|^2}
\, .
\end{equation}

The first-class constraints are $\pi_\lambda = 0$ eq.
(\ref{22}) and $G_1 = 0$ eq. (\ref{23}). Note that they are
in involution with the Hamiltonian eq. (\ref{31}). The
formulation eq. (\ref{31}) is classically equivalent to the
initial one eq. (\ref{12}). This means that, in the appropriate
gauge, the equations of motion for the physical variables in 
eq. (\ref{31}) are the same as for eq. (\ref{12}). One may
demonstrate that in the following way. Firstly, choose the
remaining constraints in eqs. (\ref{22}) and (\ref{23}),

\begin{equation}
\label{31,5}
\pi_\lambda\, =\, 0\, , \qquad |\phi|^2\lambda\, +\, {1\over 2}
\pi_a \pi_a\, +\, \partial_x \phi^a \partial_x \phi^a\, =\, 0\, ,
\end{equation}
as the gauge fixing conditions for the first class constraints
$\pi_\lambda = 0$ and $G_1 = 0$. Then, supposing that the 
corresponding
Dirac bracket eq. (\ref{24}) is constructed, one can impose the
constraints upon the Hamiltonian eq. (\ref{31}). The resulting 
expression will be identical to the Hamiltonian eq. (\ref{21}). 
Finally, one follows all the steps presented in  subsection 
\ref{subsec:two} which showed that the model described by the 
Hamiltonian eq. (\ref{21}) is classically equivalent to the one 
described by the Hamiltonian eq. (\ref{12}). Therefore, $H$ eq. 
(\ref{31}) represents correctly the model at the classical level.

Let us point out, also, that the formulation eq. (\ref{31}) 
can be used to represent the sigma-model dynamics in a
simple form. Namely, instead of (\ref{31,5}), one can now 
choose the following
gauge: $\lambda=0$ and $\phi^a\pi_a=0$, for the constraints
$\pi_\lambda=0$ and $G_1=0$. It leads to the free equations
of motion: $\partial_0\phi^a=p_a$, $\partial_0 p_a = 
\partial_i\partial_i\phi^a$ or $\Box \phi^a =0$, for the
configuration space variables. Contrary to the previous
section, they can be immediately solved in terms of the
creation and annihilation operators which obey the
brackets following from eq. (\ref{25}). It can lead to
the possible quantization of the model in the Fock-space
representation \cite{deriglazov}.

We would like to write the functional Schr\"{o}dinger
equation for $H$ eq. (\ref{31}). For this, we shall use 
the Dirac's prescription to canonically quantize 
first-class constrained systems \cite{dirac}. As we shall 
see the functional Schr\"{o}dinger equation will be the 
same as the ones derived in Subsecs. \ref{subsec:one} 
and \ref{subsec:two}.

We start by noting that the functional Schr\"{o}dinger
method described in Subsec. \ref{subsec:one} will have a
single modification in order to comply with the Dirac's
prescription to treat first-class constrained systems.
The wave-functional will have to be annihilated by the
operatorial version of the constraints besides satisfying
the functional Schr\"{o}dinger equation \cite{jackiw1,gil}.

Observing the constraint $G_1=0$ eq. (\ref{23}), we notice
that the condition that its operatorial version ($\hat{G}_1$)
annihilates the wave-functional ($\Psi$) does not result in 
any condition upon $\Psi$. It is in fact a condition upon 
the fields, since in the fields representation all the 
operators in $\hat{G}_1$ have a multiplicative application 
upon $\Psi$. One way to obtain a restriction upon $\Psi$, from
$\hat{G}_1$ is by considering the pair ($\phi^N$, $\pi_N$) as 
the corresponding non-physical variables. Then, without 
affecting the physical sector
 variables,
one can make the canonical transformation,

\begin{equation}
\label{33}
\phi^N \to - \pi_N\, \qquad \pi_N \to \phi^N\, .
\end{equation}

This transformation changes $G_1=0$ eq. (\ref{23}) and $H$ 
eq. (\ref{31}) to,

\begin{equation}
\label{34}
\tilde{G}_1\, =\, \pi_N\pi_N\, +\, \phi^i\phi^i\, -\, 
1\, =\, 0\, ,
\end{equation}
\begin{eqnarray}
\label{35}
\tilde{H}\, = & \int & dx \{ \, {1\over 2} \left[ 
\pi_i \pi_i  - \left( {\phi^i \phi^j\over \pi_N\pi_N + 
\phi^k\phi^k}\right) \pi_i \pi_j + \partial_x \phi^i 
\partial_x \phi^i + \partial_x \pi_N \partial_x \pi_N 
\right] \nonumber \\ 
& + & {1\over 2} \left[ \phi^N \phi^N + 2 \left( {\phi^i 
\phi^N \over \pi_N\pi_N + \phi^j\phi^j} \right) \pi_i 
\pi_N - \left( {\phi^N \phi^N \over \pi_N\pi_N + 
\phi^i\phi^i} \right) \pi_N \pi_N \right] \nonumber \\
& - & \lambda ( \pi_N \pi_N + \phi^i\phi^i - 1 )
+ v_\lambda \pi_\lambda\, \}\, .
\end{eqnarray}

Now, we may write the equations for the wave-functional 
$\Psi[\phi^N, \phi^i, \lambda ]$. The first two will be
obtained by demanding that the operatorial version of the
constraints $\pi_\lambda = 0$ eq. (\ref{22}) and 
$\tilde{G}_1$ eq. (\ref{34}) annihilate $\Psi$. The last
one is the functional Schr\"{o}dinger equation and will
be derived from the operatorial version of the Hamiltonian
eq. (\ref{35}) ($\hat{\tilde{H}}$). The operatorial version 
of all the above mentioned quantities will be obtained, in 
the fields representation, by the substitution of the 
momenta by -$i$ times the functional derivatives with 
respect to canonical conjugated fields eq. (\ref{14}). They 
are,

\begin{equation}
\label{36}
{\delta \Psi \over \delta \lambda}\, =\, 0\, ,
\end{equation}
\begin{equation}
\label{37}
- {\delta^2 \Psi \over \delta (\phi^N)^2}\, +\, 
(\phi^i\phi^i - 1 ) \Psi\, =\, 0\, ,
\end{equation}
\begin{eqnarray}
\label{38}
i{\partial \Psi \over \partial t}\, =\, & \int & dx 
[\, {1\over 2} \left( - {\delta^2 \Psi \over \delta 
(\phi^i)^2} +
(\phi^i \phi^j){\delta^2 \Psi \over \delta \phi^i \delta
\phi^j} + \partial_x \phi^i \partial_x \phi^i \Psi
- \partial_x {\delta \over \delta \phi^N} \partial_x 
{\delta \over \delta \phi^N} \Psi \right) \nonumber \\
& + & {1\over 2} \left( \phi^N \phi^N \Psi - 2 
(\phi^i \phi^N) {\delta^2 \Psi \over \delta \phi^i \delta
\phi^N} + (\phi^N \phi^N) {\delta^2 \Psi \over \delta 
(\phi^N)^2} \right)\, ]\, ,
\end{eqnarray}
where we have explicitly used eq. (\ref{37}) in order to 
substitute the result of the operation of the denominator 
present in $\hat{\tilde{H}}$ upon $\Psi$.

Note that the particular factor-ordering chosen in 
(\ref{36}-\ref{38}), the same one introduced in Subsec. 
\ref{subsec:one}, preserves the classical constraint algebra 
and involution of the constraints with the Hamiltonian.

Now, we may proceed to solve eqs. (\ref{36}) and (\ref{37})
in order to learn what restrictions they will impose upon 
$\Psi$. They are not difficult to solve and result, 
respectively, in,

\begin{equation}
\label{39}
\Psi[\phi^i ,\phi^N ,\lambda , t] = \Psi[\phi^i ,\phi^N , 
t]\, ,
\end{equation}
\begin{equation}
\label{40}
\Psi[\phi^i ,\phi^N , t]\, =\, \exp \left[ \int dy 
\phi^N \sqrt{\phi^i \phi^i - 1} \right] \, \Psi_{phys}
[\phi^i ,t ]\, .
\end{equation}

Finally, we must introduce $\Psi$ eq. (\ref{40}) in
the functional Schr\"{o}dinger equation (\ref{38}) to
obtain,

\begin{equation}
\label{41}
\int dx \left(\, {1\over 2}\, \tilde{g}^{ij}{\delta^2 
\Psi_{phys} \over \delta \phi^i \delta \phi^j}\, +\, 
{1\over 2}\, g_{ij}\partial_x \phi^i \partial_x \phi^j
\Psi_{phys} \, \right)\, =\, E \Psi_{phys}\, ,
\end{equation}
where $g_{ij}$ and $\tilde{g}^{ij}$ are given, respectively,
by eqs. (\ref{6}) and (\ref{9}) and we have supposed the time
dependence of $\Psi_{phys}$, given in eq. (\ref{17}). It is 
important to mention that few terms proportional to the Dirac 
delta function of the point zero ($\delta (0)$) appear in the 
derivation of eq. (\ref{41}). They contribute an infinity 
amount of energy for the system that can be removed by the 
usual regularization techniques \cite{ryder,zuber}.

Comparing eq. (\ref{41}) with eq. (\ref{18}) we notice that
they are the same.

\section{Conclusions.}
\label{sec:four}

In this work we derived the Hamiltonian formalism of the
$O(N)$ non-linear sigma model in its original version as 
a second-class constrained field theory and then as a 
first-class constrained field theory. We treated the 
model as a second-class constrained field theory, by 
two different methods: the unconstrained and the Dirac 
second-class formalisms. We showed that the Hamiltonians
for all these versions of the model are equivalents. 
Then, for a particular factor-ordering choice, 
we wrote the functional  Schr\"{o}dinger equation for 
each derived Hamiltonian. We showed that they are all 
identical which justifies our factor-ordering choice and 
opens the way for a future quantization of the model via 
the functional Schr\"{o}dinger representation.

\acknowledgments

G. Oliveira-Neto would like to thank J. Ananias Neto for helpful
discussions and FAPEMIG for the invaluable financial support.

\end{document}